%  13 pages 

% DUBNA (7-10 APRIL 1992) = LYCEN 9212 
%                ----------
% \def\aa{\vskip 0.5 true cm}
%  \font\sbx=cmssbx10
%  \font\unhb=cmdunhb10
%  \font\itdroit=cmub10
% \headline={\hss\tenrm\folio\hss}
% --------------------------------

% \input tex_inputs:maclyon 
  \magnification\magstep1 
  \baselineskip = 0.6 true cm
% \parskip=0.5 true cm
                           
  \def\sa{\vskip 0.30 true cm}
  \def\sb{\vskip 0.60 true cm}

%  \nopagenumbers
%  \voffset = 15 true mm
%  \hoffset =  8 true mm
%  \hsize = 15 true cm
%  \vsize = 22 true cm

%  \font\msym=msym10

% \font\msim=msym10

\rightline{\bf LYCEN 9212}
\rightline{03 April 1992}

\sa
\sb
\sa

\centerline {{\bf TWO-PHOTON LASER SPECTROSCOPY OF TRANSITION IONS}}

\centerline {{\bf IN CRYSTALS~: INTER-CONFIGURATIONAL TRANSITIONS}}

\sa
\sb

\centerline {M.~Daoud and M.~Kibler}

\sa

\centerline {Institut de Physique Nucl\'eaire de Lyon} 
% \hfill\break
\centerline {IN2P3-CNRS et Universit\'e Claude Bernard} 
% \hfill\break
\centerline {43 Boulevard du 11 Novembre 1918} 
% \hfill\break
\centerline {F-69622 Villeurbanne Cedex, France} 

\sa
\sa
\sb
\sb
\sb

\baselineskip = 0.55 true cm

\sa

\centerline {\bf ABSTRACT}

\sa
\sb

\noindent Symmetry adaptation techniques are applied to the 
determination of the intensity of two-photon absorption 
transitions, between Stark levels of configurations with opposite 
parities, for transition ions in finite symmetry environments. 
The equivalence between third-order mechanisms and second- plus 
first-order mechanisms is clearly established. A compact 
formula is derived for the intensity of a two-photon transition
between two levels with well-defined symmetries. 
A set of (selection) rules which 
controls the number of 
intensity parameters  for inter-configurational transitions 
is given. The formalism is illustrated with the case of an 
ion with $4f$ configuration in tetragonal symmetry. 

\sa
\sb
\sa
\sb
\sb
\sb
\sa
\sb
\sb

\noindent Invited lecture (presented by M.~Kibler) at the ``International 
Workshop on Laser Physics'', Joint Institute for Nuclear 
Research, Dubna, Russia, 7~-~10 April 1992. 
Paper published in the International 
Journal {\bf Laser Physics 2 (1992) 704-710}. 

\vfill\eject
\baselineskip = 0.5 true cm
% \vglue 2.5 true cm

\centerline {{\bf TWO-PHOTON LASER SPECTROSCOPY OF TRANSITION IONS}}

\centerline {{\bf IN CRYSTALS~: INTER-CONFIGURATIONAL TRANSITIONS}}

% \sa
  \sb

\centerline {M.~Daoud and M.~Kibler} 

\sa

\centerline {Institut de Physique Nucl\'eaire de Lyon}
% \hfill\break
\centerline {IN2P3-CNRS et Universit\'e Claude Bernard}
% \hfill\break
% \centerline {43 Boulevard du 11 Novembre 1918}
% \hfill\break
\centerline {F-69622 Villeurbanne Cedex, France} 
  \sa
% \sa
  \sb
% \sb
\baselineskip = 0.5 true cm
% \sa
\centerline {ABSTRACT}

\leftskip  = 1.4 true cm
\rightskip = 1.4 true cm

\noindent Symmetry adaptation techniques are applied to the 
determination of the intensity of two-photon absorption 
transitions, between Stark levels of configurations with opposite 
parities, for transition ions in finite symmetry environments. 
The equivalence between third order mechanisms and second- plus 
first-order mechanisms is clearly established. A compact 
formula is derived for the intensity of a two-photon transition
between two levels with well-defined symmetries. 
A set of (selection) rules which 
controls the intensity parameters  for inter-configurational transitions 
is given. The formalism is illustrated with the case of an 
ion with $4f$ configuration in tetragonal symmetry. 

\leftskip  = 0 true cm
\rightskip = 0 true cm

  \sa
  \sb
% \sa
% \sb
\baselineskip = 0.5 true cm
\noindent {\bf 1. Introduction}

The theory of two-photon absorption spectroscopy for an ion with 
a partly-filled shell embedded in a crystal has received a great 
deal of attention in recent years [1-13]. It is the aim of this 
paper to develop a formalism for $n \ell \to n' \ell'$ 
inter-configurational two-photon transitions when the two 
involved (one-electron) configurations have opposite parities. This 
problem has been already addressed by Gayen and Hamilton [7], 
Gayen {\it et al}.~[7], Leavitt [11], Makhanek {\it et al}.~[4], 
and Sztucki and Str\c ek [9]. However, little attention has been 
paid to symmetry adaptation techniques (i.e., group-theoretical 
considerations from both a qualitative and a quantitative 
viewpoint) for inter-configurational two-photon transitions. In 
this paper, we follow the general line adopted in Refs.~[12,13] 
for            intra-configurational two-photon transitions 
in order to develop a model incorporating parity violation by 
the odd components of the crystal-field interaction and 
symmetry adaptation to the chain $O(3)^* \supset G^*$. 
(The group $G^*$ denotes the double group of the point symmetry 
group $G$ of the ion site.) The model is applied to the case of 
the Ce$^{3+}$ ion in CaF$_2$ [7] and LuPO$_4$ [14].

\noindent {\bf 2. Transition element}

We begin with the (electronic) transition matrix element $M_{i \to f}$ 
for a two-photon absorption 
% transition 
between an initial state $i$ and a final 
state $f$. We work in the framework of the electric dipolar approximation
and use single-mode excitations of energy $\hbar \omega$, 
                                  wave-vector ${\vec k}$, and 
                   polarization vector ${\vec {\cal E}}$ 
for the radiation field. (For the sake of simplicity, we first 
consider that the two absorbed photons are identical.) 
Then, the matrix element $M_{i \to f}$ 
is given by [15,16] 
$$
M_{i \to f} = \sum_v {{1}\over {\Delta_v}} 
\left(f \vert \vec D. \, \vec {\cal E} \vert v \right) 
\left(v \vert \vec D. \, \vec {\cal E} \vert i \right) 
\qquad \quad {\Delta}_v = E_i - (E_v - \hbar \omega) 
\eqno (1)
$$
with evident notations. For linear polarization, the spherical 
components of $\vec {\cal E}$ are 
$$
{\cal E}_0       = \cos \theta \qquad \quad 
{\cal E}_{\pm 1} = \mp 
{1 \over {\sqrt 2}} \; \sin \theta \; \exp (\pm i \varphi) 
\eqno (2)
$$
where $(\theta, \varphi)$ are the polar angles of the unit 
polarization vector $\vec {\cal E}$ with respect to the 
crystallographic $c$-axis. For circular polarization, the 
components ${\cal E}_q$ ($q = -1,0,+1$) of $\vec {\cal E}$ are  
given by
$$
{\cal E}_{-q} = - \delta (q,+1) \quad \hbox{or} \quad 
{\cal E}_{-q} = - \delta (q,-1) 
\eqno (3)
$$
for right or left circular polarization, respectively, when the 
wave vector $\vec k$ is parallel to the $c$-axis.

The state vectors corresponding to $i$ and $f$ may be 
(partially) labelled by the irreducible representation 
classes of the group $G^*$. We thus have 
$$
\vert i ) \equiv \vert n  \ell  i \Gamma  \gamma ) \quad \qquad 
\vert f ) \equiv \vert n' \ell' f \Gamma' \gamma')
\eqno (4)
$$
where the labels $\gamma$ or $\gamma'$ may serve as 
multiplicity labels when the dimensions of the irreducible 
representation classes $\Gamma$ or $\Gamma'$ are greater than 1, 
respectively. The state vectors (4) can be expanded in terms of 
symmetry adapted state vectors as 
$$
\eqalign{
\vert n  \ell  i \Gamma  \gamma ) 
&= \sum_{j  a } \; \vert n  s \ell  j  a  \Gamma  \gamma ) 
\; c(n  \ell  j  a  \Gamma ; i) 
\cr
\vert n' \ell' f \Gamma' \gamma')
&= \sum_{j' a'} \; \vert n' s \ell' j' a' \Gamma' \gamma') 
\; c(n' \ell' j' a' \Gamma'; f) 
}
\eqno (5)
$$
with 
$$
\eqalign{
\vert n  s \ell  j  a  \Gamma  \gamma ) \; &= \; \sum_{m } \; 
\vert n  s \ell  j  m ) \; (j  m  \vert j  a  \Gamma  \gamma ) 
\cr 
\vert n' s \ell' j' a' \Gamma' \gamma') \; &= \; \sum_{m'} \; 
\vert n' s \ell' j' m') \; (j' m' \vert j' a' \Gamma' \gamma') 
}
\eqno (6)
$$
where the expansion coefficients of type 
$(JM \vert J a \Gamma \gamma)$ in (6) are reduction
coefficients to pass from the $\{ JM \}$ scheme to the 
                              $\{ J a \Gamma \gamma \}$ scheme [17]. 
The coefficients $c$ in (5) are obtainable from the 
diagonalization-optimization of the Hamiltonian $H$ describing 
the ion in its environment. (In (5) and (6), $s$ stands for 
1/2.) 

The transition matrix element $M_{i \to f}$ given by (1) and 
(4)-(6) vanishes since the initial and final states have 
opposite parities. In order to generate non-vanishing 
transition matrix elements, we must admit that the parity of 
the initial and final state vectors is not well-defined. This 
is the case if the crystal-field part of $H$ contains odd 
components either because the site symmetry is 
non-centrosymmetric or because of the vibration of the ligands. 
Therefore, we have to replace the initial and final state 
vectors, respectively, by
$$
\vert i > \> = 
\vert i ) - {1 \over \Delta} \sum_{x'} 
\vert n' \ell' x' \Gamma \gamma) 
     (n' \ell' x' \Gamma \gamma \vert O 
\vert n  \ell  i  \Gamma \gamma)
\eqno (7)
$$
and
$$
\vert f > \> = 
\vert f ) + {1 \over \Delta} \sum_x 
\vert n  \ell  x \Gamma' \gamma') 
     (n  \ell  x \Gamma' \gamma' \vert O 
\vert n' \ell' f \Gamma' \gamma')
\eqno (8)
$$
In deriving (7) and (8), we have made use of the following 
points. 

(i) We have supposed that the scrambling between 
the state vectors of opposite parities is restriced to the 
interaction, via the operator $O$, of the configurations 
$n  \ell $ and $n' \ell'$. The operator $O$ can be viewed as 
the odd part of the crystal-field interaction contained in $H$ 
and thus can be developed either as 
$$
O \, = \, \sum_k \sum_q     \, B[kq]   \, C^k_q 
\eqno (9a)
$$
in terms of standard components $C^k_q$ or as
$$
O \, = \, \sum_k \sum_{a_0} \, B[ka_0] \, C^k_{a_0 \Gamma_0 \gamma_0}
\eqno (9b)
$$ 
in terms of $G$-invariant components $C^k_{a_0 \Gamma_0 \gamma_0}$. 
(The label $\Gamma_0$ stands for the identity representation 
class of the group $G$.) The one-electron tensor operator $C^k$ in 
(9), with {\it spherical components} $C^k_q$ (adapted to the 
chain $O(3) \supset O(2)$) or {\it symmetry adapted components} 
$C^k_{a_0 \Gamma_0 \gamma_0}$ (adapted to the chain $O(3) 
\supset G$), is defined through
$$
( \ell' \Vert C^k \Vert \ell ) = (-)^{\ell'} 
[ (2 \ell' + 1) (2 \ell + 1) ]^{1 \over 2} 
\pmatrix{
\ell' & k & \ell \cr
0     & 0 & 0    \cr
}
\eqno (10)
$$
(The crystal-field parameters $B[kq]$ in (9a) correspond to 
the Rajnak-Wybourne parameters in a somewhat different form 
since we want to reserve the notation $B^k_q$ for the $q$-th 
component of a tensor $B^k$ to be defined below.) 

(ii) Furthermore, we have taken into account that, due to the 
$G$-invariance of the operator $O$, the quantum numbers $\Gamma 
\gamma$ and $\Gamma' \gamma'$ remain good quantum numbers for 
the states $i$ and $f$, respectively.

(iii) Finally, we have applied the (quasi-closure) approximation 
(cf.~Refs.~[18,19]) 
$$
\eqalign{
E(n' \ell' x' \Gamma  \gamma ) - 
E(n  \ell  i  \Gamma  \gamma ) &= E_{n' \ell'} - E_{n  \ell } \equiv 
+ \Delta \cr 
E(n  \ell  x  \Gamma' \gamma') - 
E(n' \ell' f  \Gamma' \gamma') &= E_{n  \ell } - E_{n' \ell'} \equiv 
- \Delta 
}
\eqno (11)
$$
where $\Delta = 2 \hbar \omega$, a function of the common 
energy $\hbar \omega$ of the two identical photons. 

The next approximation is to suppose that the intermediate 
states $v$ 
in (1) are mixed by the operator $O$ in a way similar to the 
one for the states $i$ and $f$ and that, in addition, they may have 
a predominant character of type $n \ell$ or $n' \ell'$. Consequently, 
we take 
$$
\vert v > \> = \vert n \ell x_v \Gamma_v \gamma_v)
- {1 \over \Delta} \sum_{{x_v}'} 
\vert n' \ell' {x_v}' \Gamma_v \gamma_v) 
     (n' \ell' {x_v}' \Gamma_v \gamma_v \vert O 
\vert n  \ell  x_v    \Gamma_v \gamma_v)
\eqno (12)
$$
or
$$
\vert v > \> = \vert n' \ell' x_v \Gamma_v \gamma_v)
+ {1 \over \Delta} \sum_{{x_v}'} 
\vert n \ell {x_v}' \Gamma_v \gamma_v) 
     (n \ell {x_v}' \Gamma_v \gamma_v \vert O 
\vert n'\ell'x_v    \Gamma_v \gamma_v)
\eqno (13)
$$
according to whether as the state $v$ has an $n \ell$ or $n' 
\ell'$ predominant character. 

The transition matrix element (1) now can be non-zero if we 
replace the initial, intermediate and final state vectors by 
the corresponding mixed state vectors. As a result, we obtain
$$
\eqalign{
- \> {1 \over 2} \> \Delta^2 \> & M(2+1)_{i \to f} = \cr
   & \sum_{x  x_v 
                  \Gamma_v \gamma_v } \; 
% ( n' \ell' f      \Gamma'  \gamma'  \vert O \vert 
(          f                        \vert O \vert 
  n  \ell  x      \Gamma'  \gamma'  ) 
( n  \ell  x      \Gamma'  \gamma'  \vert \vec D. \, \vec {\cal E} \vert 
  n' \ell' x_v    \Gamma_v \gamma_v ) 
( n' \ell' x_v    \Gamma_v \gamma_v \vert \vec D. \, \vec {\cal E} \vert 
           i                        ) \cr
% n  \ell  i      \Gamma   \gamma   ) \cr
+2& \sum_{{x_v}' x_v 
                  \Gamma_v \gamma_v} \; 
% ( n' \ell' f      \Gamma'  \gamma'  \vert \vec D. \, \vec {\cal E} \vert 
(          f                        \vert \vec D. \, \vec {\cal E} \vert 
  n  \ell  x_v    \Gamma_v \gamma_v ) 
( n  \ell  x_v    \Gamma_v \gamma_v \vert O \vert 
  n' \ell' {x_v}' \Gamma_v \gamma_v ) 
( n' \ell' {x_v}' \Gamma_v \gamma_v \vert \vec D. \, \vec {\cal E} \vert 
           i                        ) \quad (14) \cr
% n  \ell  i      \Gamma   \gamma   ) \cr
 + & \sum_{x' x_v 
                  \Gamma_v \gamma_v} \; 
% ( n' \ell' f      \Gamma'  \gamma'  \vert \vec D. \, \vec {\cal E} \vert 
(          f                        \vert \vec D. \, \vec {\cal E} \vert 
  n  \ell  x_v    \Gamma_v \gamma_v ) 
( n  \ell  x_v    \Gamma_v \gamma_v \vert \vec D. \, \vec {\cal E} \vert 
  n' \ell' x'     \Gamma   \gamma   ) 
( n' \ell' x'     \Gamma   \gamma   \vert O \vert
           i                        ) \cr  
% n  \ell  i      \Gamma   \gamma   ) \cr 
}
$$
where we have neglected the terms in $(1/\Delta)^4$ in 
$M_{i \to f}$ and taken into consideration the hermiticity 
property of the operator $O$. The notation $M(2+1)_{i \to f}$ 
for $M_{i \to f}$ indicates that the transition matrix element 
has been calculated from second-order time-dependent 
perturbation theory and first-order time-independent 
perturbation theory.

It is interesting to note that the result (14) can be obtained 
in a different way by invoking third-order mechanisms and by 
making use of approximations similar to the ones employed for deriving 
$M(2+1)_{i \to f}$. By third-order mechanisms, we mean that we 
start from the transition moment for a three-photon absorption 
transition where we replace one of the operators $\vec D. \, \vec {\cal E}$ 
by the operator $O$ with change of denominators and appropriate permutations. 
This leads to
$$
\eqalign{
M(3)_{i \to f}  = & 
\sum_{v_1 v_2} \; { {1} \over {\Delta_{12}} } \; 
( f \vert O \vert v_1 ) 
( v_1 \vert \vec D. \, \vec {\cal E} \vert v_2 ) 
( v_2 \vert \vec D. \, \vec {\cal E} \vert i ) \cr 
+ & 
\sum_{v_1 v_2} \; { {1} \over {\Delta_{12}} } \; 
( f \vert \vec D. \, \vec {\cal E} \vert v_1 ) 
( v_1 \vert O \vert v_2 ) 
( v_2 \vert \vec D. \, \vec {\cal E} \vert i ) \cr
+ &
\sum_{v_1 v_2} \; { {1} \over {\Delta_{12}} } \; 
( f \vert \vec D. \, \vec {\cal E} \vert v_1 ) 
( v_1 \vert \vec D. \, \vec {\cal E} \vert v_2 ) 
( v_2 \vert O \vert i ) \cr 
}
\eqno (15)
$$
The notation $M(3)_{i \to f}$ for $M_{i \to f}$ is a reminder that 
the calculation of the transition matrix element arises from 
third-order time-dependent perturbation theory. The state vectors 
in (15) are non-mixed state vectors. Furthermore, we assume 
in (15) that the intermediate states $v$ are of either 
$n \ell$ or $n' \ell'$ type. By using the approximation (11), we get
$$
\Delta_{12} ({\rm 1st \ sum}) = - {1 \over 2} \Delta^2 \qquad
\Delta_{12} ({\rm 2nd \ sum}) = - {1 \over 4} \Delta^2 \qquad
\Delta_{12} ({\rm 3rd \ sum}) = - {1 \over 2} \Delta^2 
\eqno (16)
$$
so that we finally arrive at the important result  
$$
M(3)_{i \to f} = M(2+1)_{i \to f}
\eqno (17)
$$ 
At this stage, it should be noted that the equivalence 
described by (17) also 
holds when the two absorbed photons are not identical. In this 
case, we have to replace 
$\Delta$ by $\hbar \omega_1 + \hbar \omega_2$ (for more 
details, see the thesis by one of the authors (M.~D.)).
 
\noindent {\bf 3. Intensities}

Equation (14) and its analogue in terms of non-identical 
photons can be worked out by using recoupling and symmetry 
adaptation techniques. In the case of two non-identical 
photons, this leads to
$$
M_{i \to f} = ( n' \ell' f \Gamma' \gamma' \vert H_{eff}    
          \vert n  \ell  i \Gamma  \gamma )   
\eqno (18)
$$
where the effective transition operator $H_{eff}$ reads
$$
H_{eff} = \sum_{hkt} M[(hk)t] 
\left( \{ \{ {\cal E}_1 {\cal E}_2 \}^h B^k \}^t \, . \, \, U^t \right) 
\eqno (19) 
$$
In eq.~(19), $\{ {\cal E}_1 {\cal E}_2 \}^h$ is the tensor product of the 
polarization 
vectors ${\vec {\cal E}}_1$ and ${\vec {\cal E}}_2$ for the two 
photons with $h = 0,1,2$. (When ${\vec {\cal E}}_1 = 
                                 {\vec {\cal E}}_2$, we have 
only         $h = 0,  2$.) Furthermore, the tensor $B^k$ in (19) 
has its spherical components $B^k_q$ given by 
$$
B^k_q = (-)^q \, B[k,-q]
\eqno (20)
$$
where the parameters $B[kq]$ are the expansion coefficients of 
the crystal-field interaction $O$ (see eq.~(9a)). Hence, $k$ 
may take the values $\vert \ell - \ell' \vert    $,    
                    $\vert \ell - \ell' \vert + 2$, $\dots$, 
                    $      \ell + \ell'          $. Finally, the 
(one-electron) tensor operator $U^t \equiv u^t$ in (19) is such that 
$$
  ( \ell  \Vert u^t \Vert \ell' ) = 1 
% ( \ell' \Vert T^t \Vert \ell  ) = - 
% ( \ell' \Vert C^1 \Vert \ell  )^2 \, ( \ell' \Vert C^k \Vert \ell)
\eqno (21)
$$
and its rank $t$ is restricted by the fact that ($\ell,\ell',t$), 
($j,j',t$), and ($h,k,t$) should form three triads. 

The parameters $M[(hk)t]$ occurring in (19) can be written as 
$$
\eqalign{
M&[(hk)t] = {1 \over \Delta} \> [(2t+1) (2h+1)]^{1 \over2} \> 
e^2 \> (n \ell \vert r \vert n' \ell')^2 \> 
( \ell \Vert C^1 \Vert \ell' )^2 \> ( \ell \Vert C^k \Vert \ell')
\cr
& \left[ 
x \> 
\left\{
\matrix{
h     & k     & t     \cr
1     & \ell  & \ell' \cr
1     & \ell' & \ell  \cr 
}
\right\}
+ y \> 
\left\{
\matrix{
h     & k     & t    \cr
\ell' & \ell  & \ell \cr
}
\right\}
\> 
\left\{
\matrix{
1     & 1     & h     \cr
\ell  & \ell  & \ell' \cr 
}
\right\} 
+ z \> 
\left\{
\matrix{
h     & k     & t     \cr
\ell  & \ell' & \ell' \cr
}
\right\}
\> 
\left\{
\matrix{
1     & 1     & h    \cr
\ell' & \ell' & \ell \cr 
}
\right\}
\right] 
}
\eqno (22)
$$
where $\Delta$ stands here for 
$\hbar \omega_1 + \hbar \omega_2 = E_{n' \ell'} - E_{n \ell}$ 
and the factors $x$, $y$, and $z$ are
$$
x = (-)^t \> [1 + (-)^h] 
\left[ {1 \over {\hbar \omega_1}} +       {1 \over {\hbar \omega_2}} \right]
\qquad 
y = (-)^t 
\left[ {1 \over {\hbar \omega_1}} + (-)^h {1 \over {\hbar \omega_2}} \right]
\qquad 
z = (-)^{t + 1} \> y 
\eqno (23)
$$
It is to be mentioned that, in the special case where the photons 
1 and 2 are identical, 
% (or have the same energy and/or the same polarization?), 
the expressions for $M[(hk)t]$ and $H_{eff}$ agree with those derived 
by Leavitt [11] from second quantization techniques and by Sztucki and 
Str\c ek [9] from ordinary Wigner-Racah calculus. (The results 
of Leavitt and of Sztucki and Str\c ek are derived in a 
spherical basis rather than in a basis adapted to the chain 
$O(3)^* \supset G^*$.)

By making use of symmetry adaptation techniques for the chain
$O(3)^* \supset G^*$, it can be shown, from eqs.~(5), (18) and 
(19), that 
$$
\eqalign{
M_{i \to f} = & - \sum_{j'a'}       \sum_{ja}       \sum_{hkt} 
                  \sum_{a''}        \sum_{\Gamma''} \sum_{\gamma''} 
                  \sum_{{\bar a}''} \sum_{a_0} 
c(j' a' \Gamma';f)^* \> c(j a \Gamma;i) \cr 
& (2 t + 1)^{1 \over 2} \> 
(s \ell  j  \Vert u^t \Vert s \ell' j' ) \> M[(hk)t] \> B[ka_0]^* 
\> \{ {\cal E}_1 {\cal E}_2 \}^h_{{\bar a}'' \Gamma'' \gamma''} \cr 
&f
\pmatrix{
t                     & h                            & k                     \cr
a'' \Gamma'' \gamma'' & {\bar a}'' \Gamma'' \gamma'' & a_0 \Gamma_0 \gamma_0 \cr
}^* \, 
 f
\pmatrix{
j                     & j'                           & t                     \cr
a   \Gamma   \gamma   & a'         \Gamma'  \gamma'  & a'' \Gamma'' \gamma'' \cr
}^* 
}
\eqno (24)
$$
where the $f$ coupling coefficients are 
defined by means of the generic formula [17]
$$
\eqalign{
f
\pmatrix{
  j_1 &   j_2 & k   \cr
\mu_1 & \mu_2 & \mu \cr
}
\; = \; \sum_{m_1 q m_2} \; & (-)^{j_1 - m_1} 
\pmatrix{
 j_1&k&j_2\cr
-m_1&q&m_2\cr
} \cr
& (j_1 m_1 | j_1 \mu_1 )^* \; 
  (k   q   | k   \mu   )   \; 
  (j_2 m_2 | j_2 \mu_2 ) 
} 
\eqno (25)
$$
with the abbreviations $\mu_i \equiv a_i \Gamma_i \gamma_i$ 
($i = 1,2$) and 
$\mu \equiv a \Gamma \gamma$. (Note that the $f$ coefficient is 
simply a symmetry adapted version of the coupling factor which 
occurs, besides  the reduced matrix element, 
in the ordinary Wigner-Eckart theorem.) In order to 
obtain eq.~(24), we have used a property of the $f$ 
coefficients which arises when 
$\mu \equiv a_0 \Gamma_0 \gamma_0$ (see Ref.~[17]).

The next step is to calculate the intensity strength 
$$
S_{  \Gamma  \to   \Gamma' } \; \equiv \; 
S_{i(\Gamma) \to f(\Gamma')} \; = \; \sum_{\gamma \gamma'} \; 
\left\vert M_{i(\Gamma \gamma) \to f(\Gamma' \gamma')} \right\vert ^2 
\eqno (26)
$$
where the sum on $\gamma$ and $\gamma'$ must be extended over 
all the Stark components of the initial and final states 
(of symmetry $\Gamma$ and $\Gamma'$), respectively. 
The calculation of $S_{  \Gamma  \to   \Gamma' }$ can be 
achieved by using~: (i) eq.~(24) twice, (ii) the factorization 
lemma for the $f$ coefficients [17], and (iii) the 
orthonormality-completeness property for the $f$ coefficients 
[17]. As a net result, we get the compact formula 
$$
S_{\Gamma \to \Gamma'} \; = \; \sum_{h,{\bar h}} 
                            \; \sum_{r,{\bar r}} \; \sum_{\Gamma''} \;
I[h {\bar h} r {\bar r} \Gamma'' ; \Gamma \Gamma'] \; 
\sum_{\gamma''} \; \left\{ 
{\cal E}_1 \, {\cal E}_2\right\}^{h}     _{r        \Gamma'' \gamma''} \; 
\left( \left\{ 
{\cal E}_1 \, {\cal E}_2\right\}^{\bar h}_{{\bar r} \Gamma'' \gamma''}\right)^* 
\eqno (27)
$$
The intensity parameters $I$ in eq.~(27) may be developed as 
$$
\eqalign{
I [ h \bar h r \bar r \Gamma'' ; \Gamma \Gamma' ] = &
[\Gamma'']^{-1} \> [\Gamma] \> 
\sum_{j'a'} \sum_{ja} \sum_{\bar j' \bar a'} \sum_{\bar j \bar a} 
\sum_{tp} \sum_{\bar t \bar p} \cr
& 
Z_{     h}(     j'      a' \Gamma' ,      j      a \Gamma , 
     h      r \Gamma'' ,      t      p \Gamma'') 
\> 
Z_{\bar h}(\bar j' \bar a' \Gamma' , \bar j \bar a \Gamma , 
\bar h \bar r \Gamma'' , \bar t \bar p \Gamma'')^* \cr
&\sum_\beta \; 
(     j'       a' \Gamma' +      t      p \Gamma'' \vert 
      j        a \beta \Gamma) \; 
(\bar j' \bar  a' \Gamma' + \bar t \bar p \Gamma'' \vert 
 \bar j  \bar  a \beta \Gamma)^*\cr
}
\eqno (28)
$$
where the coefficient 
$$
\eqalign{
Z_{     h}(     j'      a' \Gamma' ,      j      a \Gamma , 
     h      r \Gamma'' ,      t      p \Gamma'') 
= & (2j+1)^{-{1 \over 2}} \> (2t+1)^{{1 \over 2}} 
\> c(j' a' \Gamma' ; f)^* \> c(j  a  \Gamma  ; i) 
\> (s \ell  j  \Vert u^t \Vert s \ell' j' )
\cr 
& \sum_{k} \sum_{a_0} M[(hk)t] \> B[ka_0]^* \> f
\pmatrix{
t                   & h                   & k                     \cr
p \Gamma'' \gamma'' & r \Gamma'' \gamma'' & a_0 \Gamma_0 \gamma_0 \cr
}^* 
}
\eqno (29)
$$
turns out to be independent of the label $\gamma''$. The 
coefficients $( \ + \ \vert \ )$ in (28) are isoscalar factors 
for the reduction $O(3)^* \to G^*$ [17].

The intensity formula (27) has the same form as the one 
obtained in a previous work [12] for 
intra-configurational two-photon transitions. The intensity 
parameters $I$ in (27) exhibit the symmetry property
$$
I [ h \bar h r \bar r \Gamma'' ; \Gamma \Gamma' ] = 
I [ \bar h h \bar r r \Gamma'' ; \Gamma \Gamma' ]^* 
\eqno (30)
$$
which corresponds to the real character of 
$S_{\Gamma \to \Gamma'}$. Furthermore, they can be factorized 
into two similar factors when the group $G^*$ is 
multiplicity-free since the sum over $\beta$ in (28) disappears 
in this case. 

The number of {\it a priori} independent intensity parameters 
$I$ in (27) is controlled by the property (30) and the following 
selection rules. 

{\bf Rule 1}. In most cases, there is no summation in (27) over the 
multiplicity labels $r$ and $\bar r$. 

{\bf Rule 2}. The summation in (27) over $\Gamma''$ is restricted by 
the group-theoretical rules
$$
\Gamma'' \subset (     h) \qquad \quad
\Gamma'' \subset (\bar h) \qquad \quad
\Gamma'' \subset \Gamma \otimes {\Gamma'}^*
\eqno (31)
$$
where $(h)$ and $(\bar h)$ are irreducible representation classes 
of $O(3)$. Similarly, we must have 
$\Gamma'' \subset (     t)$ and 
$\Gamma'' \subset (\bar t)$ in (28) and (29). 

{\bf Rule 3}. In the cases where either 
${\vec {\cal E}_1} =   {\vec {\cal E}_2}$ or 
$\hbar \omega_1 = \hbar \omega_2$, the summation in (27) over 
$h$ and $\bar h$ is restricted to the values $0,2$ but $h$ and 
$\bar h$ can take the values $0,1,2$ when, simultaneously, 
${\vec {\cal E}_1} \ne {\vec {\cal E}_2}$ and 
$\hbar \omega_1 \ne \hbar \omega_2$. 

\noindent {\bf 4. Illustration}

We now illustrate the model developed in the present paper 
with the case of a $4f \to 5d$ transition for an ion of 
configuration $4f^1$ in tetragonal symmetry with 
$G= C_{4v}$ or $D_{2d}$. In this case, there are four possible 
transitions since $\Gamma$ and $\Gamma'$ may be equal to 
$\Gamma_6$ or $\Gamma_7$. We consider only the situation where 
the two photons are identical. Then, the possible values of $h$ 
and $\bar h$ in (27) are 0,2 and there is no summation 
on $r$ and $\bar r$ in (27). In addition, the irreducible 
representation classes $\Gamma''$ in (27) are 
$\Gamma'' = A_1, E$ for the  transitions 
$\Gamma_6 \to \Gamma_6$ or $\Gamma_7 \to \Gamma_7$ and 
$\Gamma'' = B_1, B_2, E$ for the  transitions 
$\Gamma_6 \to \Gamma_7$ or $\Gamma_7 \to \Gamma_6$. By 
developing (27), we get 
$$
\eqalign{
S_{\Gamma \to \Gamma} = 
&
I[02A_1; \Gamma \Gamma] \, 
\{ {\cal E} {\cal E} \}^0_{A_1}     \, (\{ {\cal E} {\cal E} \}^2_{A_1})^* 
+
I[20A_1; \Gamma \Gamma] \, 
(\{ {\cal E} {\cal E} \}^0_{A_1})^* \,  \{ {\cal E} {\cal E} \}^2_{A_1} 
+ \cr 
&
I[00A_1; \Gamma \Gamma] \,
                \vert \{ {\cal E} {\cal E} \}^0_{A_1       } \vert^2
+
I[22A_1; \Gamma \Gamma] \,
                \vert \{ {\cal E} {\cal E} \}^2_{A_1       } \vert^2
+
I[22E; \Gamma \Gamma] \, 
\sum_{\gamma''} \vert \{ {\cal E} {\cal E} \}^2_{E \gamma''} \vert^2
\cr
}
\eqno (32)
$$
for $\Gamma =   \Gamma' = \Gamma_6$ or $\Gamma_7$ and 
$$
S_{\Gamma \to \Gamma'} = 
I[22B_1; \Gamma \Gamma'] \> 
                \vert \{ {\cal E} {\cal E} \}^2_{B_1       } \vert^2
+
I[22B_2; \Gamma \Gamma'] \> 
                \vert \{ {\cal E} {\cal E} \}^2_{B_2       } \vert^2
+
I[22E  ; \Gamma \Gamma'] \> 
\sum_{\gamma''} \vert \{ {\cal E} {\cal E} \}^2_{E \gamma''} \vert^2
\eqno (33)
$$
for $\Gamma \ne \Gamma' = \Gamma_6$ or $\Gamma_7$. The 
polarization factors in (32) and (33) can be calculated 
in an $O(3) \supset C_{\infty v} \supset C_{4v} \supset C_{2v}$ 
basis to be 
$$
\eqalign{
\{{\cal E} {\cal E}\}^0_{A_1}&={-1 \over \sqrt{3}} 
  \quad {\rm or} \quad 0 \cr
\{{\cal E} {\cal E}\}^2_{A_1}&={1 \over \sqrt{6}} ({3 \cos^2 \theta - 1})  
  \quad {\rm or} \quad 0 \cr
\{{\cal E} {\cal E}\}^2_{B_1}&={1 \over \sqrt{2}} \sin^2 \theta \cos 2 \varphi 
  \quad {\rm or} \quad +  {1 \over \sqrt{2}} \cr
\{{\cal E} {\cal E}\}^2_{B_2}&={i \over \sqrt{2}} \sin^2 \theta \sin 2 \varphi 
  \quad {\rm or} \quad \pm{1 \over \sqrt{2}} \cr
\sum_{\gamma''} \vert \{{\cal E} {\cal E}\}^2_{E \gamma''} \vert^2
&={1 \over 2} \sin^2 2 \theta 
  \quad {\rm or} \quad 0 \cr
}
\eqno (34)
$$
according to as the polarization is linear or circular, 
respectively. In the detail, equations (32) and (33) lead to
(for linear polarization) 
$$
\eqalign{
S_{\Gamma_6 \to \Gamma_6} \; & = \; 
a  + b \, \pi_1 + c \, \pi_1^2 + d \, \pi_2 \cr 
S_{\Gamma_7 \to \Gamma_7} \; & = \;
a' + b'\, \pi_1 + c'\, \pi_1^2 + d'\, \pi_2 \cr 
S_{\Gamma_6 \to \Gamma_7} \; & = \;
f \, \pi_2 + h \, \pi_4 + i \, \pi_5        \cr
S_{\Gamma_7 \to \Gamma_6} \; &= \;
f'\, \pi_2 + h'\, \pi_4 + i'\, \pi_5        \cr
}\eqno (35)
$$
where the angular functions $\pi_i$ ($i=1,2,4,5$) are defined by
$$
\pi_1 = 3 \; \cos^2 \theta \; - \; 1        \quad \qquad 
\pi_2 = \sin^2 2 \theta                     \quad \qquad 
\pi_4 = \sin^4   \theta \; \cos^2 2 \varphi \quad \qquad 
\pi_5 = \sin^4   \theta \; \sin^2 2 \varphi 
\eqno (36)
$$
and the various parameters $a, \cdots, i$ and $a', \cdots, i'$ 
read
$$
\eqalign{
  a \; &= \; (1/3)         \;           I[00A_1;\Gamma_6\Gamma_6]   \cr
  b \; &= \; -(\sqrt{2}/3) \; {\rm Re}[ I[02A_1;\Gamma_6\Gamma_6] ] \cr
  c \; &= \; (1/6)         \;           I[22A_1;\Gamma_6\Gamma_6]   \cr
  d \; &= \; (1/2)         \;           I[22E  ;\Gamma_6\Gamma_6]   \cr
  f \; &= \; (1/2)         \;           I[22E  ;\Gamma_6\Gamma_7]   \cr
  h \; &= \; (1/2)         \;           I[22B_1;\Gamma_6\Gamma_7]   \cr
  i \; &= \; (1/2)         \;           I[22B_2;\Gamma_6\Gamma_7]   \cr
}
\qquad \quad
\eqalign{
  a' \; &= \; (1/3)        \;           I[00A_1;\Gamma_7\Gamma_7]   \cr
  b' \; &= \; -(\sqrt{2}/3)\; {\rm Re}[ I[02A_1;\Gamma_7\Gamma_7] ] \cr
  c' \; &= \; (1/6)        \;           I[22A_1;\Gamma_7\Gamma_7]   \cr
  d' \; &= \; (1/2)        \;           I[22E  ;\Gamma_7\Gamma_7]   \cr
  f' \; &= \; (1/2)        \;           I[22E  ;\Gamma_7\Gamma_6]   \cr
  h' \; &= \; (1/2)        \;           I[22B_1;\Gamma_7\Gamma_6]   \cr
  i' \; &= \; (1/2)        \;           I[22B_2;\Gamma_7\Gamma_6]   \cr
}
\eqno (37)
$$
in terms of the intensity parameters $I$. The latter 
parameters are calculable from (28) and (29) together with 
(22). (Equation (28) shows that the parameters $I$ can be 
expressed here as the product of two similar factors since there is 
no internal multiplicity label $\beta$ for tetragonal symmetry.)

4.1. Ce$^{3+}$ in CaF$_2$

Two-photon experiments for the ion Ce$^{3+}$ (of configuration 
$4f^1$) in CaF$_2$ (with local site symmetry $G=C_{4v}$) have 
been reported in Ref.~[7]. The polarization 
dependence of the inter-configurational transition 
$^2F_{5/2}(\Gamma_7) \to 5d(\Gamma_7)$ (between the ground 
state of the $4f^1$ configuration and           the ground 
state of the $5d^1$ configuration) has 
been interpreted by Gayen and Hamilton [7] on the basis of the 
group-theoretical formalism from Bader and Gold [3] and by 
Makhanek {\it et al}.~[4] on a more quantitative basis. 

From equations (35) and (36), the intensity of the 
$^2F_{5/2}(\Gamma_7) \to 5d(\Gamma_7)$ transition can be 
rewritten as
$$
S_{\Gamma_7 \to \Gamma_7} = 
A + B \sin^2 \theta + C \sin^2 2 \theta 
\eqno (38)
$$
with
$$
\eqalign{
A & = (1/3) \,                 I[00A_1; \Gamma_7 \Gamma_7] 
    + (2/3) \,                 I[22A_1; \Gamma_7 \Gamma_7] 
    - 2(\sqrt{2}/3)\,{\rm Re} [I[02A_1; \Gamma_7 \Gamma_7]] \cr 
B & =-(1/2) \,                 I[22A_1; \Gamma_7 \Gamma_7] 
    + \sqrt{2}     \,{\rm Re} [I[02A_1; \Gamma_7 \Gamma_7]] \cr 
C & =-(3/8) \,                 I[22A_1; \Gamma_7 \Gamma_7] 
    + (1/2) \,                 I[22E  ; \Gamma_7 \Gamma_7]] \cr 
}
\eqno (39)
$$
The F$^-$ charge compensator in CaF$_2$ has an equal probability of 
going into an interstitial site along the [100], [010] or [001] 
directions. Therefore, the intensity strength of the 
$^2F_{5/2}(\Gamma_7) \to 5d(\Gamma_7)$ transition has to be averaged as
$$
S_{\Gamma_7 \to \Gamma_7} = {1 \over 3} \left( 
S([100])_{\Gamma_7 \to \Gamma_7} +
S([010])_{\Gamma_7 \to \Gamma_7} +
S([001])_{\Gamma_7 \to \Gamma_7} \right)
\eqno (40)
$$
Equations (38) and (40) yield 
$$
\eqalign{
S_{\Gamma_7 \to \Gamma_7} & = A + (2/3) B + (2/3) C \sin^2 2 \theta \cr
S_{\Gamma_7 \to \Gamma_7} & = A + (1/3) B + (2/3) C 
+ (1/3) B \sin^2 \theta - (1/3) C \sin^2 2 \theta \cr
S_{\Gamma_7 \to \Gamma_7} & = A + (2/3) B + (2/3) C \cr
}
\eqno (41)
$$
for (${\vec k} \Vert [100]$ ${\vec {\cal E}} \Vert [010]$), 
    (${\vec k} \Vert [110]$ ${\vec {\cal E}} \Vert [001]$), and 
    (${\vec k} \Vert [111]$ ${\vec {\cal E}} \Vert [1{\bar 1}0]$), 
respectively, which correspond to 
the three experimental situations considered by Gayen and 
Hamilton [7]. The intensity formulas (41) formally exhibit the same 
$\theta$-dependence than those obtained in Ref.~[7]. Equations 
(41), where the intensity parameters $A$, $B$ and $C$ can be 
{\it a priori} calculated from the formalism developed in 
section 3, constitutes a justification of the purely 
group-theoretical approach used in Ref.~[7]. (Let us 
remember that the approach used in Ref.~[7] is based on 
group-theoretical selection rules for intra-configurational 
transitions.) Finally, it 
should be noted that the scalar terms (corresponding to $h$ or 
$\bar h = 0$ in (27)), described in (41) by the 
intensity parameters $I[00A_1; \Gamma_7 \Gamma_7]$ and 
                     $I[02A_1; \Gamma_7 \Gamma_7]$, do not 
occur in the treatment of Ref.~[4]. 

4.2. Ce$^{3+}$ in LuPO$_4$

Inter-configurational two-photon transitions for Ce$^{3+}$ in 
LuPO$_4$ (site symmetry $G=D_{2d}$) have been 
observed by Piehler [14]. The transitions investigated are of the 
type $^2F_{5/2}(\Gamma_6) \to 5d$ (between the ground state 
of the $4f^1$ configuration and the Stark components 
of the $5d^1$ configuration) with linear polarization. Piehler 
has obtained a first transition 
(at $2 \hbar \omega =       30460$ cm$^{-1}$) 
whose intensity vanishes when the two (identical) 
photons are polarized along the $z$-axis and a second one 
(at $2 \hbar \omega \approx 40000$ cm$^{-1}$)
whose intensity does not vanish for polarization along the 
$z$-axis. 

The symmetry species of the $5d$ final states reached in Ref.~[14] 
can be easily determined from (35) and (36). Indeed, from (35) and 
(36), the intensities of the transitions 
$^2F_{5/2}(\Gamma_6) \to 5d (\Gamma_6$ or $\Gamma_7)$ 
can be seen to be
$$
\eqalign{
S_{\Gamma_6 \to \Gamma_6} & = 
A' + B' \sin^2 \theta              + C' \sin^2 2 \theta \cr 
S_{\Gamma_6 \to \Gamma_7} & = 
A'' \sin^4 \theta \cos^2 2 \varphi + 
B'' \sin^4 \theta \sin^2 2 \varphi + C'' \sin^2 2 \theta \cr
}\eqno (42)
$$
where 
$$
\eqalign{
A' & = (1/3) \,                  I[00A_1; \Gamma_6 \Gamma_6] 
     + (2/3) \,                  I[22A_1; \Gamma_6 \Gamma_6] 
     - 2(\sqrt{2}/3)\, {\rm Re} [I[02A_1; \Gamma_6 \Gamma_6]] \cr 
B' & =-(1/2) \,                  I[22A_1; \Gamma_6 \Gamma_6] 
     + \sqrt{2}     \, {\rm Re} [I[02A_1; \Gamma_6 \Gamma_6]] \cr 
C' & =-(3/8) \,                  I[22A_1; \Gamma_6 \Gamma_6] 
     + (1/2) \,                  I[22E  ; \Gamma_6 \Gamma_6]] \cr 
A'' & = (1/2)\,                  I[22B_1; \Gamma_6 \Gamma_7]  
\quad 
B''   = (1/2)\,                  I[22B_2; \Gamma_6 \Gamma_7]  
\quad 
C''   = (1/2)\,                  I[22E  ; \Gamma_6 \Gamma_7]  \cr 
}
\eqno (43)
$$
Thus, equation (42) shows that only the intensity strength 
$S_{\Gamma_6 \to \Gamma_7}$ vanishes when the electric field is 
polarized along the $z$-axis (i.e., $\theta = 0$). As a 
consequence, the symmetry of the final states is $\Gamma_7$ for 
the first transition 
(at $30460$ cm$^{-1}$) 
and $\Gamma_6$ for the second transition
(at $40000$ cm$^{-1}$). 
This result is in accordance with the one obtained by 
Piehler [14] on the basis of selection rules derived 
from (qualitative) group-theoretical considerations. 
Here again, it is to be mentioned that the intensities (42) 
are calculable from {\it ab initio} principles 
once the intensity parameters (43) are known. 

\noindent {\bf 5. Concluding remarks}

The main result of this paper is formula (27) which gives the 
polarization dependence of an inter-configurational two-photon 
transition between Stark components arising from the 
configurations $n \ell$ and $n' \ell'$ of opposite 
parities. The obtained formula bears the same form as the 
corresponding one for intra-configurational two-photon 
transitions. 

The intensity parameters $I$ in (27) are, likewise in the case 
of intra-configurational two-photon transitions, 
model-dependent. More precisely, they depend on the 
wave-functions for the initial and final states and on 
the odd crystal-field parameters. They also incorporate 
the information coming from the involved symmetry group. 
They can be calculated in an 
{\it ab initio} way or may be considered as phenomenological 
parameters. 

Finally, we would like to mention that the model developed in 
the present work applies to the case of 
$n \ell^N \to n \ell^{N-1} n' \ell'$ 
two-photon transitions with $\ell + \ell'$ odd. (The passage 
from one-electron configurations to $N$-electron configurations 
does not affect the symmetry considerations.)

\noindent {\bf Acknowledgments}

The authors are grateful to B.~Jacquier for communicating the 
thesis by D.~Piehler (cf.~Ref.~[14]). Thanks are due to 
G.W.~Burdick, J.C.~G\^acon, and B.~Jacquier for discussions. The 
results of this paper were presented at the 
``International Workshop on Laser Physics'' (JINR, Dubna, Russia, 
April 1992)~; one of the authors (M.~K.) wishes to thank the 
Organizing Committee for inviting him to give a talk at this 
workshop. 

\vfill\eject 
\baselineskip 0.52 true cm

\noindent {\bf References} 

\item{[1]} J.D. Axe, Jr., Phys. Rev. 136 (1964) A42.

\item{[2]} M. Inoue and Y. Toyozawa, J. Phys. Soc. Japan 
20 (1965) 363. 

\item{[3]} T.R. Bader and A. Gold, Phys. Rev. 171 (1968) 997.

\item{[4]} P.A. Apanasevich, R.I. Gintoft, V.S. Korolkov, A.G.
Makhanek and G.A. Skripko, Phys. Status Solidi (b) 58 (1973) 
745~; A.G. Makhanek and G.A. Skripko, Phys. Status Solidi (a) 53 (1979) 
243~; A.G. Makhanek, V.S. Korolkov 
and L.A. Yuguryan, Phys. Status Solidi (b) 149 (1988) 231.

\item{[5]} B.R. Judd and D.R. Pooler, J. Phys. C : Solid State Phys.
15 (1982) 591.

\item{[6]} M.C. Downer and A. Bivas, Phys. Rev. B~28 (1983) 
3677~; M.C. Downer, in~: Laser Spectroscopy of Solids II, 
Ed. W.M. Yen (Springer, Heidelberg, 1989)~; G.W. Burdick and M.C. 
Downer, to be published. 

\item{[7]} S.K. Gayen and D.S. Hamilton,  Phys. Rev. B~28 (1983) 
3706~;  S.K. Gayen, D.S. Hamilton and R.H. Bartram, 
Phys. Rev. B~34 (1986) 7517. 

\item{[8]} M.F. Reid and F.S. Richardson, Phys.~Rev.~B~29 (1984) 
2830. 

\item{[9]} J. Sztucki and W. Str\c ek, Phys. Rev. B~34 (1986) 
3120~; Chem. Phys. Lett. 125 (1986) 520~; Chem. Phys. 143 (1990) 347.

\item{[10]} K. Jankowski and L. Smentek-Mielczarek 
Molec.~Phys.~60 (1987) 1211~;    L. Smentek-Mielczarek 
and B.A. Hess, Jr., Phys.~Rev.~B 36 (1987) 1811. 

\item{[11]} R.C. Leavitt, Phys. Rev. B~35 (1987) 9271. 

\item{[12]} M. Kibler and J.C. G\^acon, Croat. Chem. Acta 62
(1989) 783~; M. Kibler, in~: Symmetry and Structural Properties 
of Condensed Matter, Eds.  W. Florek, T. Lulek and M. Mucha 
(World, Singapore, 1991)~; 
M. Kibler and M. Daoud, in~: Proc.~V Workshop on 
Symmetry Methods in Physics, Obninsk, USSR, July 1991, in the 
press~; M. Kibler, in~: Proc.~IInd International School on 
Excited States of Transition Elements, Karpacz, Poland, September 
1991, World Scientific, Singapore, in the press. 

\item{[13]} J.C. G\^acon, J.F. Marcerou, M. Bouazaoui, B. Jacquier
and M. Kibler, Phys. Rev. B~40 (1989) 2070~; J.C. G\^acon, B. Jacquier, 
J.F. Marcerou, M. Bouazaoui and M. Kibler, J. Lumin. 45 (1990) 
162~; J.C. G\^acon, M. Bouazaoui, B. Jacquier, M. Kibler, L.A. 
Boatner and M.M. Abraham, Eur. J. Solid State Inorg. Chem. 28 (1991) 
113~; J. Sztucki, M. Daoud and M. Kibler, Phys. Rev. 
B~45 (1992) 2023. 

\item{[14]} D. Piehler, Ph.~D.~ thesis, University of 
California, Berkeley, California (1990). 

\item{[15]} R. Loudon, The Quantum Theory of Light 
(Clarendon, Oxford, 1973).

\item{[16]} C. Cohen-Tannoudji, J. Dupont-Roc et G. 
Grynberg, Processus d'interaction entre photons et atomes 
(InterEditions et Editions du CNRS, Paris, 1988).

\item{[17]} M. Kibler, C.R. Acad. Sc. (Paris) B 268 (1969) 
1221~; M.R. Kibler and P.A.M. Guichon, Int. J. Quantum 
Chem. 10 (1976) 87~;
M.R. Kibler and G. Grenet, Int. J. Quantum Chem. 11 (1977) 359~;
M.R. Kibler, in~: Recent Advances in Group 
Theory and Their Application to Spectroscopy, Ed. J.C. Donini 
(Plenum Press, N.Y., 1979)~; Int. J. Quantum Chem. 23 (1983) 
115~; M. Kibler and G. Grenet, Studies in 
Crystal-Field Theory (Report LYCEN/8656, IPNL, Lyon, 1986).

\item{[18]} B.R. Judd, Phys. Rev. 127 (1962) 750.

\item{[19]} G.S. Ofelt, J. Chem. Phys. 37 (1962) 511. 

\bye